\documentstyle[aps]{revtex}

\input epsf.tex 
\begin{document}
\title{Dark energy and the Boomerang data}
\author{Luca Amendola}
\address{Osservatorio Astronomico di Roma, \\
Viale del Parco Mellini 84, \\
00136 Roma, Italy\\
{\it amendola@oarhp1.rm.astro.it}}
\date{\today }
\maketitle

\begin{abstract}
The recent high-quality Boomerang data allow to test many competing
cosmological models. Here I present a seven-parameter likelihood analysis of
dark energy models with exponential potential and explicit coupling to dark
matter. Such a model is conformally equivalent to a scalar field with
non-minimal coupling to gravity. So far, the constraints on a dark energy -
dark matter coupling were extremely weak. The Boomerang data constrain the
dimensionless coupling $\beta $ to be smaller than 0.1, an order of
magnitude better than previous limits. In terms of the constant $\xi $ of non-minimally
coupled theories, this amounts to $\xi <0.01.$

On the other hand, Boomerang has not enough sensitivity to put constraints
on the potential slope.
\end{abstract}

\section{Introduction}

In the last few years, a new reference model of structure formation has
emerged, one in which 70\% or so of the total matter content of the universe
is in a form of dark energy. The main evidence for such a component is the
supernovae Ia observations of Ref. \cite{per} and \cite{rie}, which can be
explained by assuming an accelerated expansion due either to a cosmological
constant or to a new matter component with equation of state 
\begin{equation}
p=(w-1)\rho
\end{equation}
with $w\in (0,0.6)$ \cite{per}\cite{wang2} . This new component can also be
modeled as a light scalar field with a potential that allows for a
potential-dominated epoch \cite{fri} \cite{cal} as, e.g., an exponential
potential or an inverse power law .

The evidence for the new component, sometimes denoted dark energy or
''quintessence'' \cite{cal}, has been reinforced by the recent Boomerang CMB
observations that show a preference for a flat universe \cite{deb}\cite{lan}
. In fact, since a CDM density of $\Omega _{c}\in (0.2,0.4)$ is in agreement
with a host of independent observations, from cluster abundance to cluster
X-ray temperature, lensing, velocity fields etc., we can conclude that a
conspicuous fraction of the total energy density has to be unclustered (or
weakly clustered) and with negative pressure \cite{dod}, although the direct
constraint on the amount of dark energy from CMB alone is weak

The nature of this extra component is so far completely unknown. The
simplest explanation, the cosmological constant, runs against the argument
that, from a dimensional point of view, its value should be more than a
hundred orders of magnitude smaller than expected, and tuned with
astonishing precision in order to become dominant just today. The hypothesis
of some fundamental scalar field has at least the advantage that one may
hope to build a theory that explains the coincidence as a consequence of
some fundamental principle. For instance, the inclusion of a coupling
between dark matter and dark energy, as will be done in this paper, might
explain why the two energy densities are comparable. Another motivation for
considering a scalar field lies in the fact that it is premature to be too
specific about the dark energy properties: a general scalar field includes
as a limiting case the cosmological constant, but allows also to investigate
less extreme effective equations of state.

In the same spirit of generality, I investigate the effect of an additional
degree of freedom, represented by the coupling of the scalar field to
ordinary matter. Such a coupling has been proposed and investigated in refs. 
\cite{wan}\cite{wet95}\cite{car}\cite{bil}\cite{hol}\cite{ame99}, and is
equivalent, up to a conformal rescaling, to the classical Brans-Dicke
coupling to gravity (for gravity coupling in the context of quintessence see
refs. \cite{uza} \cite{chi}\cite{che}\cite{bac}\cite{far}). If the dark
energy is coupled to dark matter alone (and not to baryons), the present
constraints on the coupling are rather weak \cite{dam}\cite{cas}\cite{hol};
as a consequence, in \cite{ame00b} it was shown that the effects on the CMB
turn out to be at the level of detectability already with the present data
set.

Aim of this paper is to use the most recent CMB data, the Boomerang power
spectrum \cite{lan}, along with the COBE data (as reduced in \cite{bon}), to
further constrain the coupled dark energy model. As particular cases, we
will derive constraints on the pure cosmological constant model and on the
uncoupled dark energy. A comparison of Boomerang with a different model of
uncoupled dark energy is in \cite{bra}.

In all the calculations of this paper a flat universe is assumed. This is of
course a severe limitation, but the number of free parameters is already
large enough to be at the limits of computational capabilities. Moreover,
the hypothesis of quintessence has always been formulated in the context of
flat models, in order to be consistent with the inflationary expectations.

While the work was almost completed the results of the Maxima experiment 
\cite{han}\cite{bal}\cite{jaf} have been published. They seem to confirm the
results of Boomerang, but have not been included in the present paper.

\section{Coupled dark energy}

The model of coupled quintessence or dark energy has already been studied in
detail in \cite{ame00b} . Here I limit myself to a summary of its properties.

Consider three components, a scalar field $\phi $ , baryons and CDM,
described by the energy-momentum tensors $T_{\mu \nu (\phi )},$ $T_{\mu \nu
(b)}$and $T_{\mu \nu (c)}$, respectively. General covariance requires the
conservation of their sum, so that it is possible to consider a coupling
such that 
\begin{eqnarray}
T_{\nu (\phi );\mu }^{\mu } &=&\left( C_{c}T_{(c)}+C_{b}T_{(b)}\right) \phi
_{;\nu },  \nonumber \\
T_{\nu (c);\mu }^{\mu } &=&-C_{c}T_{(c)}\phi _{;\nu }.  \nonumber \\
T_{\nu (b);\mu }^{\mu } &=&-C_{b}T_{(b)}\phi _{;\nu }.  \label{coup1}
\end{eqnarray}
where $C_{c,b}$ are the coupling constants. This particular coupling is indeed
obtained by conformally transforming a non-minimally coupled gravity theory or,
following \cite{dam}, by the Lagrangian (in units  $G=c=1$) 
\begin{equation}
L_{tot}=-\frac{R}{2\kappa ^{2}}+\frac{1}{2}\phi _{;\mu }\phi ^{;\mu }-U(\phi
)+L_{c}(e^{2C_{c}\phi }g_{\mu \nu })+L_{b}(e^{2C_{b}\phi }g_{\mu \nu })
\end{equation}
where $\kappa ^{2}=8\pi $ and where $L_{c,b}$ denote the Lagrangian of the
dark matter and baryonic fields. Notice that our constant $C_{c}$
corresponds to the invisible coupling constant $\beta _{I}$ of \cite{dam}.
The radiation field (subscript $\gamma $) remains uncoupled, since $%
T_{(\gamma )}=0.$ We derive the background equations in the flat conformal
FRW metric $ds^{2}=a^{2}(-d\tau ^{2}+\delta _{ij}dx^{i}dx^{j})$ assuming the
exponential potential 
\begin{equation}
U=Ae^{s\phi }
\end{equation}
as proposed e.g. in \cite{rat}\cite{wet95}\cite{fer}. As anticipated, we
will couple the dark energy scalar field to the dark matter only, putting $%
C_{b}=0$. We call this choice dark-dark coupling. Generalizing \cite{cop} we
introduce the following variables (putting $H=\dot{a}/a$) 
\begin{equation}
x=\frac{\kappa }{H}\frac{\dot{\phi}}{\sqrt{6}},\quad y=\frac{\kappa a}{H}%
\sqrt{\frac{U}{3}},\quad z=\frac{\kappa a}{H}\sqrt{\frac{\rho _{\gamma }}{3},%
}\quad v=\frac{\kappa a}{H}\sqrt{\frac{\rho _{b}}{3},}\quad 
\end{equation}
Adopting the $e$-folding time $\alpha =\log a$ we can write the field,
baryon and radiation conservation equation as a system in the four variables 
$x,y,z,v$ that depends on the parameters $\mu ,\beta $ 
\begin{eqnarray}
x^{\prime } &=&\left( \frac{z^{\prime }}{z}-1\right) x-\mu y^{2}+\beta
(1-x^{2}-y^{2}-z^{2}-v^{2}),  \nonumber \\
y^{\prime } &=&\mu xy+y\left( 2+\frac{z^{\prime }}{z}\right) ,  \nonumber \\
z^{\prime } &=&-\frac{z}{2}\left( 1-3x^{2}+3y^{2}-z^{2}\right) ,  \nonumber
\\
v^{\prime } &=&-\frac{v}{2}\left( -3x^{2}+3y^{2}-z^{2}\right) 
\end{eqnarray}
where the prime denotes derivation with respect to $\alpha $, and where $%
\beta =C_{c}\sqrt{\frac{3}{2\kappa ^{2}}},\quad \mu =s\sqrt{\frac{3}{2\kappa
^{2}}}.$ The \ coupling constant $\beta $ can therefore be regarded as the
ratio of the dark energy coupling strength to the gravitational strength.
The dark-dark coupling $\beta $ is unconstrained by local experiments or by $%
\dot{G}/G$ measurements \cite{dam}. Constraints can be derived only by
global effects on the cosmological dynamics: for instance, from the
requirement of a universe older than 10 Gyr Ref. \cite{dam} found 
\begin{equation}
|\beta |<1
\end{equation}
In the following we show that Boomerang puts a much more stringent
constraint on the dark-dark coupling.

Neglecting the baryons, the system has a transient attractor (properly
speaking, a saddle point) at the point $x_{0}=2\beta /3,y_{0}=0,z_{0}=0$ on
which the scale factor expands as 
$
a\sim t^{6/(9+4\beta ^{2})}
$
(here $t$ is the ordinary time defined as $d\tau =a(t)dt$), that is, slower
than in a ordinary MDE. During this stage, first found in Ref. \cite{nar},
and denoted $\phi $MDE in \cite{ame00b}, the scalar field kinetic energy and
the dark matter have a constant density ratio. When the universe leaves the
saddle, it reaches a global attractor if $\mu <3$, at the point $x_{0}=-\mu
/3,y_{0}=(1-\mu ^{2}/9)^{1/2},z_{0}=0$ and the universe expansion follows
the law 
$
a\sim t^{3/\mu ^{2}}
$
mimicking a perfect fluid with equation of state $p=(w_{\phi }-1)\rho $,
with $w_{\phi }=2\mu ^{2}/9$ . The expansion is accelerated only if $\mu < 
\sqrt{3}$. As long as the baryon component is small with respect to the CDM,
the phase space of the system remains qualitatively the same.

It can be noticed that the $\phi $MDE depends only on $\beta$, while the
subsequent $\phi$-dominated epoch depends only on $\mu$. Therefore, as shown
in \cite{ame00}, the possibility to set constraints on the coupling $\beta$
depends on the existence of the $\phi $MDE. This epoch is actually quite
more general than appears from the discussion above: it exists infact for
any potential such that $U(\phi)$ and $dU/d\phi$ go asymptotically to zero,
as for instance when $U$ is an inverse power law.

The coupled quintessence with exponential potential contains the case of
pure cosmological constant ($\mu =0,\beta =0$), of uncoupled quintessence ($%
\beta =0$) and is asymptotically equivalent to a perfect fluid with a
constant equation of state $w_{\phi }=2\mu ^{2}/9$. The model is conformally
equivalent to a non-minimal gravity theory with Lagrangian term $-\frac{1}{2}%
\xi \psi ^{2}R$ and potential $V(\psi )=\lambda \psi ^{n}$ with $\xi =2\beta
^{2}/3$ and $n=4+\mu /\beta $ \cite{wet95}\cite{ame00}\cite{ame00b}.
Therefore, a bound on $\beta $ amounts to constraining the non-minimal
gravity coupling $\xi $. Notice that in the case we study here, in which the
dark energy couples differently to baryons and to dark matter, there are two
conformally related metrics in the Jordan frame (the frame in which gravity
is coupled to the scalar field): the constant $\xi $ couples the scalar
field to the Ricci scalar expressed in the metric in which the dark matter,
rather than the baryons, follow the geodetics (see \cite{dam}).

The perturbation equations, derived in \cite{ame00}, are integrated by the
use of a purposedly modified CMBFAST code \cite{sel}. In addition to the
scalar field, baryons, CDM and radiation, the code includes also massless
neutrinos. I choose adiabatic initial conditions, as suggested by inflation.
The initial conditions for the background equations are found for each set
of parameters by trial and error so as to give today $\Omega _{b},\Omega
_{c} $ and $H_{0}$ as requested. This procedure is so time consuming with
respect to the code without coupled scalar field that the assumption of flat
space and a further reduction in the parameter space explained below turned
out to be necessary.

\section{Likelihood analysis}

Our theoretical model depends on two scalar field parameters and four
cosmological parameters: 
\begin{equation}
\beta ,\mu ,n_{s},h,\Omega _{b},\Omega _{c}
\end{equation}
The remaining input parameters requested by the CMBFAST code are set as
follows: $T_{cmb}=2.726K,$ $Y_{He}=0.24,N_{\nu }=3.04,\tau _{c}=0.$ The
latter quantity specifies the optical depth to Thomson scattering, and is a
measure of reionization. In the analysis of \cite{lan} this parameter was
also included in the general likelihood and, in the flat case, was found to
be compatible with zero at slightly more than 1$\sigma $ . Moreover, in
ref. \cite{lan} it is found that fixing $\tau =0$ has only a minor effect on
the other parameters. Therefore here, to further reduce the parameter space,
I assume $\tau _{c}$ to vanish. Two other approximations with respect to 
\cite{lan} have been necessary: first, I did not include the
cross-correlation between bandpowers because it is not available. Second, an
offset log-normal approximation to the band-power likelihood has been
advocated by \cite{bon} and adopted by \cite{lan}, but the $x$-quantity
necessary for its evaluation is not available. Since the offset log-normal
reduces to a log-normal in the limit of small noise I evaluated the
log-normal likelihood 
\begin{equation}
-2\log L(\alpha _{j})=\sum_{i}\frac{\left[ Z_{\ell ,t}(\ell _{i};\alpha
_{j})-Z_{\ell ,d}(\ell _{i})\right] ^{2}}{\sigma _{\ell }^{2}}
\end{equation}
where $Z_{\ell }\equiv \log \hat{C}_{\ell }$, the subscripts $t$ and $d$
refer to the theoretical quantity and to the real data, $\hat{C}_{\ell }$
are the spectra binned over some interval of multipoles centered on $\ell
_{i}$, $\sigma _{\ell }^{2}$ are the experimental errors on $Z_{\ell ,d}$,
and the parameters are denoted collectively as $\alpha _{j}$. A 10\%
calibration error is added to the experimental errors \cite{lan}. I also
evaluated for comparison a likelihood gaussian in the variables $\hat{C}%
_{\ell }$, and found that the results do not change appreciably up to 2$%
\sigma $ from the peak.

Among the parameters that refer to the scalar field, we have already shown
in \cite{ame00b} that $\mu $ has a negligible effect on the background
solution at $z\gg 1$, since the equivalence time does not depend on it and,
although $\mu $ sets the speed of the present accelerated expansion, this
has only a minor effect on the perturbation spectrum at decoupling. This
will be confirmed below. Also, in \cite{ame00b} we have shown that the
dynamics of the system is insensitive to the sign of $\beta $, since both
the $\phi $MDE and the final accelerated epoch do not depend on it. We will
consider only $\beta \ge 0$.

In order to compare with the Boomerang analysis I assume uniform priors as
in \cite{lan}, with the parameters confined in the range $\beta \in (0,0.16),
$ $\quad \mu \in (0,2.1),\quad $\ $n_{s}\in (0.7,1.3),$ $h\in (0.45,0.9),$ $%
\quad \Omega _{b}\in (0.01,0.2),\quad $\ $\Omega _{c}\in (0.1,0.9),$ with
the further weak big-bang nucleosynthesis (BBN) constraint $\Omega
_{b}h^{2}<0.05$ (and of course the condition $\Omega _{c}+\Omega _{b}\leq 1$) .
I evaluate the likelihood also using the most up-to-date BBN limit
\cite{tyt}\cite{oli}, assuming a gaussian prior for $\Omega _{b}h^{2}$ with
mean $0.019$ and 1$\sigma$ error $0.002$ (I refer to this as the  strong BBN
prior).  The range of $\mu $ includes all the values for which there is
acceleration at the present. A further parameter, the logarithm of the absolute
normalization of the $C_{\ell }$ spectrum, is integrated out analytically with
uniform prior. The same age constraints ($>10$ Gyr) used in \cite{lan} is
adopted here. Note that here we do not have the problem of near-degeneracy of
parameters that is considered in ref. \cite{lan} . In fact, the degeneracy that
exists for those values of $\Omega _{c},\Omega _{b},h,\mu $ and $\beta $ that
give an identical angular-diameter distance to the decoupling surface is
completely removed by the combined constraints of flatness, BBN and the allowed
range of $h$ and $\mu $. We also checked that the limits on $n_{s}$ and $\beta $
are broad enough to contain the bulk of likelihood.

A grid of $\sim 10,000$ multipole CMB spectra $C_{\ell }$ is used as a
database over which I interpolate to produce the likelihood function. Since
both the COBE and the Boomerang data are in fact binned over intervals of
multipoles, I average in the same bandpowers the theoretical spectra for a
correct comparison. Three cases will be distinguished: pure $\Lambda $ ($\mu
=\beta =0$); dark energy ( $\beta =0$); coupled dark energy. In Fig. 1 I
report examples of multipole spectra obtained varying $\mu $ and $\beta $
and fixing the other parameters: $\left\{ {n_{s},\Omega _{b},\Omega _{c},h}%
\right\} =\left\{ {\ 1,0.05,0.3,0.7}\right\} $. The quantity plotted is
actually $(\ell (\ell +1)C_{\ell }/2\pi )^{1/2}\mu $K as customary. The
strong decrease in amplitude of the acoustic peaks as $\beta $ gets larger
than 0.1 depends on the fact that for these values the onset of the $\phi $%
MDE occurs before the decoupling. During the $\phi $MDE the fluctuations
smaller than the horizon grow less than during MDE, so that the fluctuations
on the acoustic scales are depressed relative to the larger scales.
Moreover, the addition of the integrated Sachs-Wolfe effect at small
multipoles decreases the normalization of the intrinsic fluctuations at
decoupling \cite{bac}\cite{ame00b}.

In Fig. 2 ( panels $a-d$, solid lines) I plot the likelihood function for
the parameters $n_{s},h,$ $\Omega _{b},\Omega _{c}$ (marginalizing over all
the other parameters) in the case of pure $\Lambda $ and weak BBN prior.
They are reasonably well in agreement with the analysis of \cite{lan} (their
case P10). The means and variances for the pure $\Lambda $ model are 
\begin{equation}
n_{s}=0.96\pm 0.06,\quad h=0.73\pm 0.09,\quad \Omega _{b}=0.056\pm
0.016,\quad \Omega _{c}=0.5\pm 0.2,
\end{equation}
while the best estimates (peaks) are $n_{s}=0.975,h=0.7,\Omega
_{b}=0.05,\Omega _{c}=0.3$ in agreement with \cite{lan} (notice that Ref. 
\cite{lan} quotes the likelihood maximum for clear detections, e.g. for $
\Omega _{b}$ and $n_{s}$, and the mean in the other cases). For the case of
uncoupled dark energy, I plot in Fig. 2 the likelihood functions in short
dashed lines, plus in the panel $e$ the new parameter $\mu $. As expected,
there is almost complete degeneracy in the direction of $\mu $, and the
results of pure $\Lambda $ remain very similar. If the Boomerang errorbars
were reduced to one third, the likelihood would begin to show some
preference for higher values of $\mu $, as shown in the same panel $e$
(dotted line). The problem of a high baryon content from Boomerang (see e.g.
ref.  \cite{bur}) is not alleviated by this model of dark energy.

The likelihood functions for the coupled dark energy case are plotted in
Fig. 2 as long-dashed lines. Due to the degeneracy along $\mu $ we can
simplify the analysis fixing this parameter to any value in the relevant
range; we put $\mu =0.25$. It turns out that the previous results are quite
robust also with respect to the coupling, except for a shift of $n_{s}$
toward smaller values ($n_{s}=0.88\pm 0.06$). In the panel $f$ I plot the
coupling $\beta $ : at the 96.8\% c.l. we obtain 
\begin{equation}
|\beta |<0.1  \label{lim}
\end{equation}

Assuming the strong BBN prior we obtain the likelihood functions shown as
dot-dashed lines in the same Fig. 2 (shown only for the coupled dark energy
case for clarity; the other cases are similar). As expected, smaller $\Omega
_{b}$ and smaller $h$ than before are now acceptable. Also, $\Omega _{c}$
moves to smaller values, in order not to decrease excessively  the first
peak. This effect, along with the decrease of $n_{s}$ in panel $b$, is
observed also in \cite{lan} when the strong BBN is imposed. The likelihood
for the exponential slope $\mu $ (not shown) becomes quite flatter with the
strong BBN.  Interestingly, $\beta $ now peaks around 0.05, because this
value raises somewhat the first peak, compensating for the low baryon
content. The limit (\ref{lim}) remains valid.

\section{Conclusions}

The most interesting conclusion that can be drawn from our analysis is that
the coupling $\beta $ between dark matter and dark energy has to be smaller
than 0.1, an order of magnitude better than previously \cite{dam},
independently of the BBN constraint. Although I derived this limit in the
particular case of an exponential potential, it extends to a much larger
class of theories since the $\phi $MDE is dominated by the kinetic, rather
than potential, energy of the scalar field, as will be shown in another
paper,

In terms of the coupling constant $\xi $ of non-minimal gravity theories we
get roughly $\xi <0.01$ at the 96.8\% c.l. .

The likelihood of the other cosmological parameters is robust with respect
to the addition of the coupling.

The potential parameter $\mu$, which sets the effective equation of state of
the dark energy at the present, is not well constrained by the present CMB,
since its effect is very recent on the cosmological time scale.

It is of course possible to add further constraints from large-scale
structure, supernovae Ia and from other CMB experiments. In \cite{ame00b} a
constraint similar to (\ref{lim}) was found from cluster abundance, for a
particular choice of the other parameters. Future data, especially
high-resolution, high-coverage experiments on CMB, have the potential to
strenghten the limit by at least another order of magnitude.

\bigskip


\newpage

\begin{figure}[tbp]
\epsfysize 8in
\epsfbox{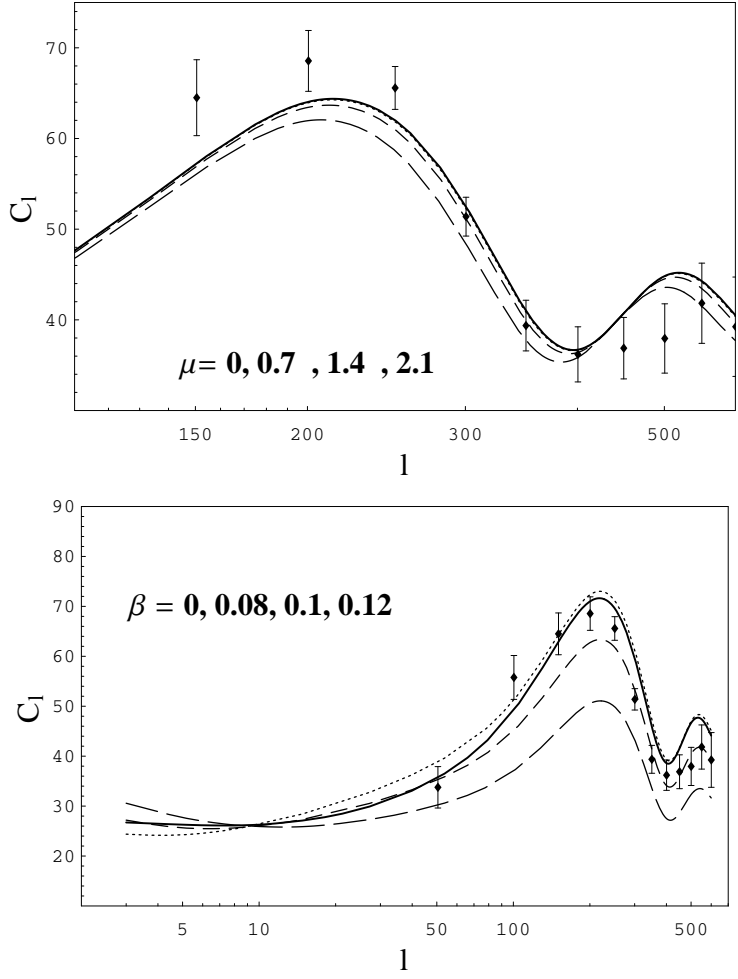}
\caption{ Top panel: spectra $C^*_\ell\equiv (\ell(\ell+1)C_\ell /2\protect%
\pi)^{1/2} \protect\mu$K of uncoupled dark energy for several values of $%
\protect\mu$ (solid: $\protect\mu=0 $; dots: $\protect\mu=0.7$; short
dashes: $\protect\mu=1.4$; long dashes: $\protect\mu=2.1$). Bottom panel:
spectra $C^*_\ell$ of coupled dark energy for several values of $\protect%
\beta$ (solid: $\protect\beta=0$; dots: $\protect\beta=0.08$; short dashes: $%
\protect\beta=0.1$; long dashes: $\protect\beta=0.12$)}
\end{figure}

\newpage

\begin{figure}[tbp]
\epsfysize 8in
\epsfbox{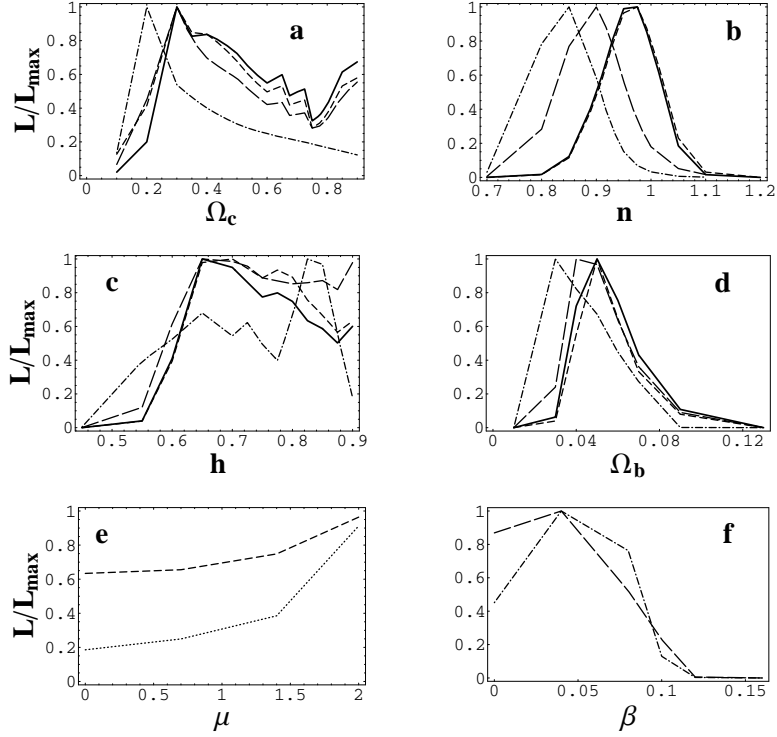}
\caption{ Marginalized likelihood functions. Solid lines: pure $\Lambda $
model. Short-dashed lines: dark energy. Long-dashed lines: coupled dark
energy. Dot-dashed lines: coupled dark energy imposing the strong BBN prior. In panel $e$ the dotted line represents the likelihood for $\protect%
\mu $ if the Boomerang errorbars were reduced to one third.}
\end{figure}

\end{document}